\documentclass[12pt]{iopart}

\usepackage{harvard}

\usepackage{graphicx}

\newcommand{\pun}[1]{\mbox{\rm\,#1}} 

\newcommand{\logg}{\ensuremath{\log g}}

\newcommand{\moh}{\ensuremath{[\mathrm{M/H}]}}

\newcommand{\Teff}{\ensuremath{T_{\mathrm{eff}}}}

\newcommand{\CIFIST}{{CIFIST}}
\newcommand{\ion}[2]{#1\,{\sc#2}}

\begin{document}

\title[Extremely metal-poor stars from the SDSS]{Extremely metal-poor stars
  from the SDSS}

\author{H.-G. Ludwig$^{1,2}$, P. Bonifacio$^{1,2}$, E. Caffau$^1$, N.T. Behara$^{1,2}$,\\
  J.I. Gonz{\'a}lez Hern{\'a}ndez$^{1,2}$, and L. Sbordone$^{1,2}$}

\address{$^1$Observatoire de Paris-Meudon, GEPI, 92195 Meudon Cedex, France}
\address{$^2$CIFIST Marie Curie Excellence Team, Observatoire de Paris-Meudon} 

\ead{Hans.Ludwig@obspm.fr}

\begin{abstract}
  We give a progress report about the activities within the \CIFIST\ Team
  related to the search for extremely metal-poor stars in the Sloan Digital
  Sky Survey's spectroscopic catalog. So far the search has provided 25
  candidates with metallicities around or smaller $-3$. For 15 candidates high
  resolution spectroscopy with UVES at the VLT has confirmed their extremely
  metal-poor status. Work is under way to extend the search to the SDSS's
  photometric catalog by augmenting the SDSS photometry, and by gauging the
  capabilities of X-shooter when going to significantly fainter targets.
\end{abstract}


\section{Introduction}

Extremely metal-poor (EMP) stars with $\moh <-3$ are exceedingly rare objects
but provide crucial information about the first generation of stars and the
chemical make-up of the early Universe. The \CIFIST\ (Cosmological Impact of
the FIrst STars) project is an EU funded Marie Curie Excellence Team hosted by
Paris Observatory with the mission of searching for and analyzing EMP stars.
The following is a progress report of activities within the \CIFIST\ Team
related to using the Sloan Digital Sky Survey (SDSS) as novel source of EMP
candidates for high-resolution spectroscopic abundance studies of EMP stars.
This includes the SDSS follow-up survey, the Sloan Extension for Galactic
Understanding and Exploration (SEGUE).

Traditionally, the HK~survey \cite{Beers+al85}, and Hamburg-ESO survey (HES)
\cite{Wisotzki+al96,Christlieb+al08} are the main sources of candidates in
searches for metal-poor stars. Both surveys are fairly deep (HK $B<15.5$, HES
$B<17.5$) objective prism, low resolution spectroscopic surveys which provided
about 10\,000 candidates each. Follow-up medium resolution ($R\approx 2000$)
spectroscopy was used for confirming EMP candidates, having typically provided
yields of a few percent of confirmed EMP stars.

Data release 6 (DR6) of the SDSS \cite{Adelman+al08} provides photometric and
spectroscopic data for about 10\,000 square degrees of the sky at mostly
northern Galactic latitudes. The major part of the data consists of photometry
in the five SDSS filters ($ugriz$) for $287\times 10^6$ unique objects. The
smaller spectroscopic part consists of medium resolution ($R\approx 2000$)
spectra of $218\times 10^3$ stars earlier than type M, as well as
$895\times 10^3$ spectra of galaxies and quasars. The spectral coverage is
380--920\pun{nm}. The spectral coverage and resolution is sufficient to
identify EMP stars with high confidence making the SDSS an interesting new
source for EMP targets.

\section{Metal-poor stars from SDSS spectra}

From the SDSS catalog we selected objects identified by the SDSS pipeline as
stars, and for which spectroscopic as well as photometric information was
available. We restricted the search to magnitudes $g<17$ ($g$ roughly
corresponds to Johnson $V$) allowing to potentially obtain high-resolution
spectra with 10\pun{m} class telescopes in a reasonable amount of time. We
further used the $g-z$ color as effective temperature (\Teff) indicator to
narrow down the selection to stars with spectral type between F and early G.
We derived a purely theoretical relation between color and effective
temperature from model atmospheres obtaining ($g$ and $z$ in magnitudes)
\begin{equation*}
\Teff\left[\mathrm{K}\right] = 7126.7 - 2844.2\,(g-z) + 666.80\,(g-z)^2 - 11.724\,(g-z)^3.
\end{equation*}
The rather high {\Teff}s of our targets made it likely that we would mainly
select dwarfs.  This was done in view of later abundance work, but also to ease
the task of analyzing the SDSS spectra by assuming a typical surface gravity
of $\logg=4.0$ (cgs). For estimating the metallicities we adapted an automatic
abundance analysis code \cite{Bonifacio+Caffau03} based on fitting of spectral
lines. At extremely low metallicities and the resolution of the SDSS spectra,
the \ion{Ca}{II}~K line is essentially the only remaining metal abundance
indicator, and was the prime target in our line fitting procedure. The
selection of stars from DR6 based on photometric indices provided about
34\,000 candidates whose metallicities were subsequently estimated with the
fitting code using the photometric temperatures and assuming $\logg=4.0$.

\begin{figure}
\centering
\resizebox{0.6\hsize}{!}{\includegraphics[clip=true]{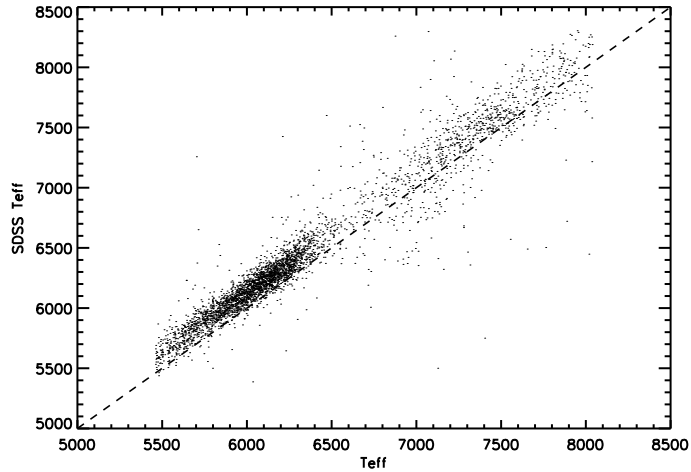}}
\caption{Comparison between effective temperatures from the SDSS catalog to
  ours from $g-z$ colors. We find on average 150\pun{K} cooler temperatures.}
\label{f:sdssteff}
\end{figure}

\begin{figure}
\centering
\resizebox{0.6\hsize}{!}{\includegraphics[clip=true]{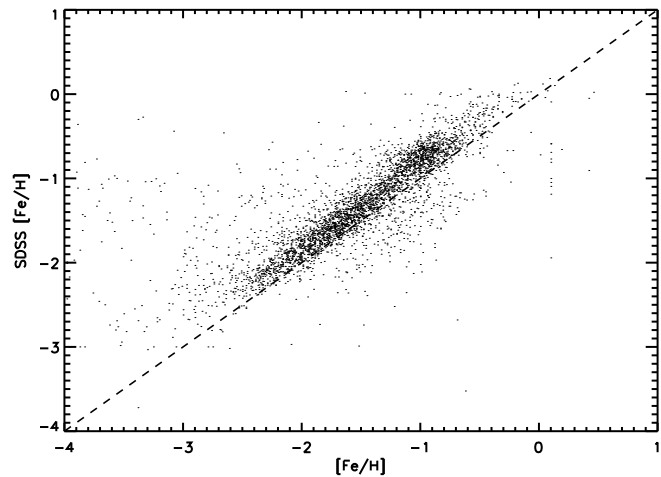}}
\caption{Comparison of metallicities from the SDSS catalog to ours
  obtained with the automatic line fitting code. We find on average
  0.2\pun{dex} lower metallicities.}
\label{f:sdssmoh}
\end{figure}

Figure~\ref{f:sdssteff} shows that our \Teff-estimates are closely correlated
with the estimates provided by the SDSS catalog albeit in our case shifted to
about 150\pun{K} lower temperatures. Similarly, Fig.~\ref{f:sdssmoh} shows a
good correlation of the metallicity estimates, however, again biased towards
lower values by about 0.2\pun{dex} in our case. To some extent the lower
metallicities are expected due to our lower \Teff-estimates. At very low
metallicities there is an extended ``halo'' of largely discrepant objects.
Visual inspection of the spectra has shown that they are commonly objects
misclassified by the SDSS pipeline, like white dwarfs or accretion disk
objects with non-stellar spectral characteristics. While one might wonder
about the remaining systematics in the stellar parameter estimation for our
purpose of identifying EMP candidates the achieved metallicity precision was
sufficient.

\begin{figure}
\centering
\resizebox{0.6\hsize}{!}{\includegraphics[clip=true]{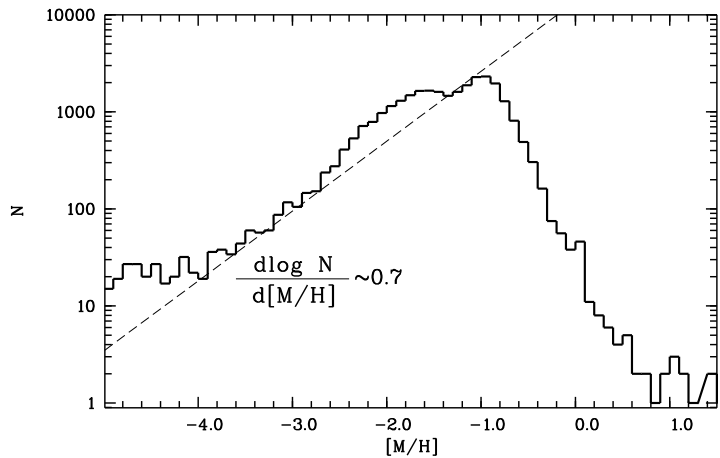}}
\caption{Metallicity distribution of 34\,000 photometrically selected F-
  to mid G-type stars. In the metallicity range $-4.0 < \moh < -2.5$ the
  selection follows a power law with exponent of about 0.7. The tail at lowest
  metallicities is an artifact due to misclassified objects.}
\label{f:mohhist}
\end{figure}

Figure~\ref{f:mohhist} shows the metallicity distribution obtained in our
selection procedure. During the course of the SDSS project, spectra of stars
have been mainly taken for calibration purposes with the tendency to pick blue
objects. Hence, there was a bias towards metal-poor objects. $ugriz$ colors do
not allow to distinguish metallicities below about $-2$. Arguably,
\cite{Carollo+al07} the distribution at lower metallicities is thus free from
selection effects, and the it reflects the intrinsic stellar distribution.
Anyways, we did not attempt to investigate this issue further, and just remark
that our distribution can be fitted by a power law with an index of about 0.7
in the range $-4.0 < \moh < -2.5$. At the lowest metallicities the
distribution is unreliable due to the strong contamination by misclassified
objects.

\section{High resolution follow-up spectroscopy}

\begin{figure}
\centering
\resizebox{0.6\hsize}{!}{\includegraphics[clip=true]{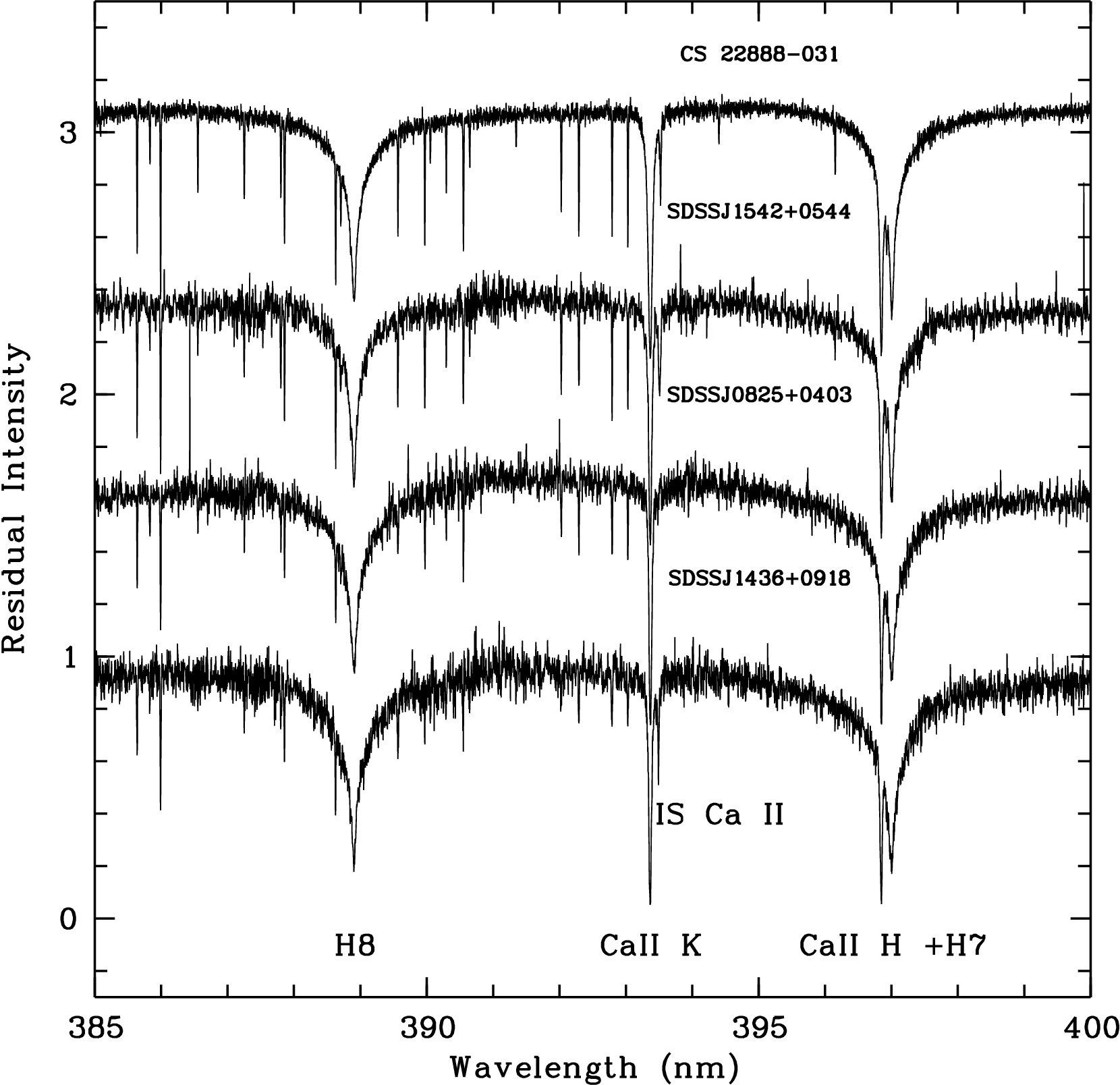}}
\caption{Comparison of UVES spectra of three EMP stars extracted from the SDSS
with the known EMP dwarf CS~22888-031 ([Fe/H]=-3.3).}
\label{f:uvesspectra}
\end{figure}

We obtained so far in total 6 nights of observing time at the VLT with UVES
for 25 EMP candidates. For 15 objects the observations are completed, and we
obtained spectra with a resolution of $R\approx 21\,000$ . The abundance
analysis is not completed yet so that only qualitative results can be
presented here. Figure~\ref{f:uvesspectra} shows a comparison of of three of
our SDSS stars in comparison to the metal-poor dwarf CS~22888-031 which has a
metallicity of [Fe/H]=-3.3. All spectra have a similar resolution, and
all stars share about the same effective temperature and gravity. Most of
the visible features which are not labeled are Fe lines. It is obvious that
our EMP candidates are at least as metal-poor as CS~22888-031. Up to now our
selection procedure has proven to be very effective, and no object with
metallicity noticeably higher than expected was found in the high resolution
spectroscopic follow-up.

In our sample of EMP stars from the SDSS carbon rich objects are quite common.
Already expected from apparently strong G-bands in the SDSS spectra, their
high carbon content has also been always confirmed by the high resolution
spectra.

\section{Tapping into the SDSS photometry}

\begin{figure}
\centering
\resizebox{0.6\hsize}{!}{\includegraphics[clip=true]{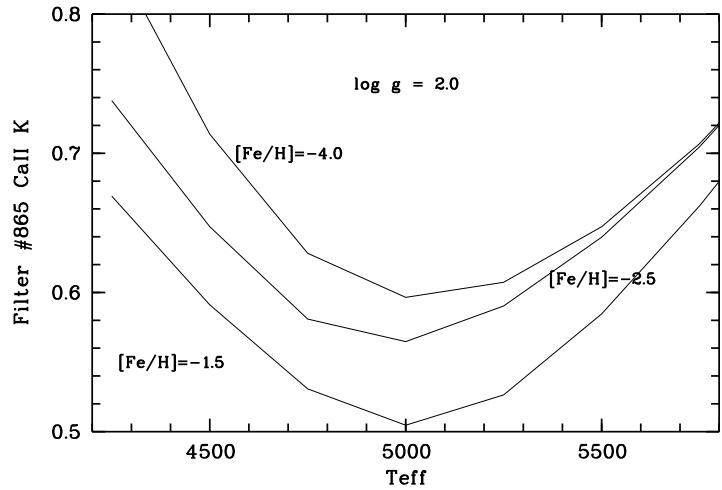}}
\caption{Response of the \ion{Ca}{II}~K filter for giants.}
\label{f:caiifiltergiants}
\end{figure}

\begin{figure}
\centering
\resizebox{0.6\hsize}{!}{\includegraphics[clip=true]{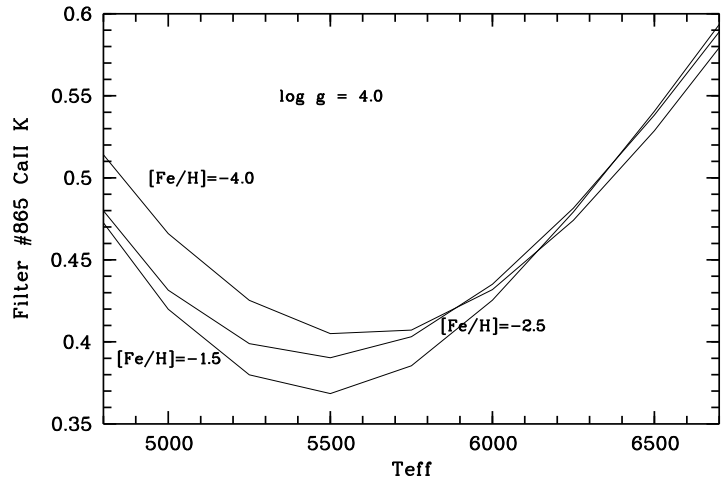}}
\caption{Response of the \ion{Ca}{II}~K filter for dwarfs.}
\label{f:caiifilterdwarfs}
\end{figure}

Unfortunately, already the rather modest number of EMP stars found so far in
the SDSS spectroscopy starts to exhaust the reservoir of solid EMP candidates.
One would wish to access the substantially larger reservoir of photometric
data.  As stated previously the SDSS photometry as such does not allow to
segregate EMP stars from moderately metal-poor objects.  To overcome this
limitation, in a pilot study we try to augment the SDSS photometry by
narrow-band photometry centered on the Ca~K line. We utilized the available
filter number~865 on the ESO 2.2\pun{m} telescope equipped with the Wide Field
Imager. Figures~\ref{f:caiifiltergiants} and~\ref{f:caiifilterdwarfs} show the
theoretical filter response curves for gravities typical for giants and
dwarfs, respectively.  The filter provides a good discrimination to low
metallicities, however, not in the F-type temperature range but for cooler
temperatures. In the filter we have imaged 10 square degrees of the sky which
has been covered by the SDSS, and are in the process to extract the
photometry. If workable, this procedure might open up a large sample of K-type
stars for the search for EMP objects. We expect the technique not to be as
effective as making use of SDSS-like spectra for identifying candidates, but
nevertheless comparable to the yields previously obtained from the HK~survey
and HES.

\section{Going deeper with X-shooter}

Soon the X-shooter spectrograph at the VLT will see first light. X-shooter
provides a wide spectral coverage with $R\approx 7000$ between the UV and
500\pun{nm}, $R\approx 12000$ from 500 to 1000\pun{nm}. This kind of
resolution is not ideal but nevertheless sufficient for some abundance work. The
outstanding efficiency of X-shooter allows to go to noticeably fainter targets
than with UVES. We obtained two nights of Guaranteed Time Observations for
testing the capabilities of X-shooter for observing EMP stars. We selected 10
candidates from the SDSS in the range $18.3<g<19.4$ of which a subset
(depending on semester) will be observed. It will be exciting to see what this
machine can do for finding the most metal-poor stars in the local Universe.

\ack 
This work was made possible by financial support from EU grant
MEXT-CT-2004-014265 (CIFIST) which is gratefully acknowledged.

\section*{References}
\bibliographystyle{jphysicsB}
\bibliography{ludwig}

\end{document}